\documentclass[
aip,
jcp,%
amsmath,amssymb,
reprint,%
]{revtex4-1}

\usepackage{color}

\newcommand{\sub}[1]{_{\mbox{\scriptsize {#1}}}}
\def\siml{\hspace{1ex} ^{<} \hspace{-2.5mm}_{\sim} \hspace{1ex}}
\def\simg{\hspace{1ex} ^{>} \hspace{-2.5mm}_{\sim} \hspace{1ex}}

\usepackage{graphicx}
\usepackage{dcolumn}
\usepackage{bm}

\begin{document}


\title[Free energy of cluster formation and a new scaling for nucleation]
{Free energy of  cluster formation  
 and a new scaling relation for the nucleation rate}

\author{Kyoko K. Tanaka}
\affiliation{Institute of Low Temperature Science, Hokkaido University, Sapporo 060-0819, Japan} 

\author{J\"urg Diemand}
\affiliation{Institute for Computational Science, University of Z\"urich, 8057 Z\"urich, Switzerland}

\author{Raymond Ang\'elil}
\affiliation{Institute for Computational Science, University of Z\"urich, 8057 Z\"urich, Switzerland}

\author{Hidekazu Tanaka}
\affiliation{Institute of Low Temperature Science, Hokkaido University, Sapporo 060-0819, Japan} 

\date{\today}

\begin{abstract}

Recent very large molecular dynamics simulations of homogeneous
nucleation with $(1-8) \cdot 10^9$ Lennard-Jones atoms [Diemand et
al. J. Chem. Phys. {\bf 139}, 074309 (2013)] allow us to accurately
determine the formation free energy of clusters over a wide range of
cluster sizes.  This is now possible because such large simulations
allow for very precise measurements of the cluster size distribution
in the steady state nucleation regime.  The peaks of the free energy
curves give critical cluster sizes, which agree well with independent
estimates based on the nucleation theorem.  Using these results, we
derive an analytical formula and a new scaling relation for nucleation
rates: $\ln J' / \eta$ is scaled by $\ln S / \eta$, where the
supersaturation ratio is $S$, $\eta$ is the dimensionless surface
energy, and $J'$ is a dimensionless nucleation rate.  This relation
can be derived using the free energy of cluster formation at
equilibrium which corresponds to the surface energy required to form
the vapor-liquid interface.  At low temperatures (below the triple
point), we find that the surface energy divided by that of the
classical nucleation theory does not depend on temperature, which
leads to the scaling relation and implies a constant, positive Tolman
length equal to half of the mean inter-particle separation in the
liquid phase.
\end{abstract}

\maketitle

\section{Introduction}

The nucleation process of supersaturated vapors into liquids (or
solids) has been studied for a long time, however, there is still a
serious gap in our understanding.  The classical nucleation theory
(CNT)\cite{Volmer1926,Becker1935,Feder1966} is a very widely used
model for describing nucleation and provides the nucleation rates as a
function of temperature, supersaturation ratio, and macroscopic
surface tension of a condensed phase.  However, several studies have
found that the CNT fails to explain the nucleation rates observed in
experiments$^{4-15}$.  For example, the error is the order of
$10^{11-20}$ for argon \cite{Iland2007,Sinha2010}.  In addition to
laboratory experiments, numerical simulations of molecular dynamics
(MD) or Monte Carlo (MC) simulations showed that the nucleation rates
obtained by numerical simulations are significantly different from
predictions by the CNT$^{16-38}$.  Until now several modifications to
the CNT were proposed. It was also noted that several nucleation rate
data sets exhibited empirical temperature scalings
\cite{Hale1986,Hale1992,Hale2005,Hale2010}.  Although there have been
significant advances in the theoretical models, a quantitatively
reliable theoretical model does not yet exist.

Recently, Diemand et al. \cite{Diemand2013} presented large-scale
molecular dynamics (MD) simulations of homogeneous vapor-to-liquid
nucleation of $(1 - 8) \times 10^9$ Lennard-Jones atoms, covering up
to 1.2 $\mu$s ($5.6 \times 10^7$ steps).  The simulations cover a wide
range of temperatures and supersaturation ratios. This study measured
various quantities such as nucleation rates, critical cluster sizes,
and sticking probabilities of vapor molecules, and it was successful
in quantitatively reproducing argon nucleation rates at the same
pressures, supersaturations and temperatures as in the SSN (Supersonic
Nozzle Nucleation) argon experiment \cite{Sinha2010}. Here we use
these MD results to determine the free energies of cluster formation
(Sec. \ref{sec:free-energy}) and their scaling
(Sec. \ref{sec:new-scaling}), which is expected to be of use in the
construction of a high-precision nucleation model.

\section{Empirical scaling relations}

\begin{figure}
\includegraphics[height=.33\textheight]{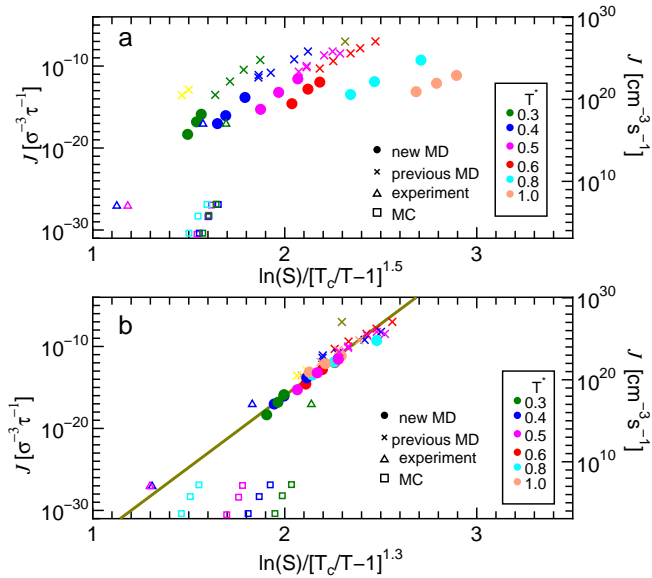}
\caption{Nucleation rates obtained by the MD simulations with LJ
 molecules and the experimental results for argon as a function of (a)
 $\ln S/(T\sub{c}/T-1)^{1.5}$ and (b) $\ln S/(T\sub{c}/T-1)^{1.3}$.
 The results for various supersaturation ratios $S$ and temperatures
 $T^*$ ($= kT/\epsilon$ in the Boltzmann constant $k$ and the depth of
 the LJ potential $\epsilon$) obtained by the large-scale MD
 simulations\cite{Diemand2013} and the previous ones
 \cite{Yasuoka1998, Wedekind2007, Tanaka2011} are shown by the filled
 circles and the crosses, respectively.  The results for MC
 simulations\cite{Hale2010} are shown with square markers, where the
 temperatures are $T^*=0.7, 0.5, 0.419,$ and 0.335.  The triangles
 show the experimental results for argon \cite{Iland2007,Sinha2010}.
 We adopt $T\sub{c}=1.312 \epsilon /k$ (or 151~K) in the simulations
 (or experiments).  In (b) the fitting function (solid line) for $J$
 $[\sigma^{-3} \tau^{-1}]$ is given by $ \log J=17.5 \ln
 S/(Tc/T-1)^{1.3}-51$.  }\label{scale}
\end{figure}

Hale and Thomason \cite{Hale2010} suggested that the nucleation rate
$J$ obtained by MC simulations using LJ molecules was scaled by $\ln
S/(T\sub{c}/T-1)^{1.5}$ over a range of
$J=(10^4-10^7)$cm$^{-3}$~s$^{-1}$ which corresponds to $(10^{-30} -
10^{-27}) \sigma^{-3} \tau^{-1}$, where $T$, $T\sub{c}$, $\sigma$, and
$\tau$ are the temperature, critical temperature, a parameter of
length ($=3.405$ \AA), and a time unit ($=2.16$~ps).
Figure~\ref{scale} shows that nucleation rates obtained by the MD and
MC simulations for LJ molecules and experimental results for argon as
a function of $\ln S/(T\sub{c}/T-1)^{1.5}$ and $\ln
S/(T\sub{c}/T-1)^{1.3}$.  The scaling by $\ln S/(T\sub{c}/T-1)^{1.5}$
works for MC simulations over a limited range, however, the nucleation
rates obtained by all MD simulations and some experiments are rather
scaled by $\ln S/(T\sub{c}/T-1)^{1.3}$.  The fitting function is $
\log J=17.5 \ln S/(Tc/T-1)^{1.3}-51$.  This linear, empirical scaling
relation seems to work well over a surprisingly wide range of
nucleation rates, $J=(10^{-30}-10^{-5}) \sigma^{-3} \tau^{-1}$ for the
MD data and the NPC (Nucleation Pulse Chamber) experiment
\cite{Iland2007}, but not for the MC simulations.  Interestingly, a
different scaling relation, $\ln S/(Tc /T -1)^{3/2}$ has been found
from experimental nucleation rates for several different substances
such as water\cite{Hale2005}, toluene\cite{Schmitt1983}, and
nonane\cite{Adams1984}.  Our results suggests the scaling relation
depends on the substance type.

However, linear empirical scaling relations contradict one of the most basic, general
expectations from nucleation theory: according to the nucleation
theorem, the size of the critical cluster $i^*$ is determined by the
derivative $d (\ln J) / d (\ln S)$ \cite{Kalikmanov,Diemand2013}.
These empirical scalings therefore imply a constant critical cluster
size $i^*$ at each temperature over a wide range in $J$.  The
corresponding free energy functions would need to peak at exactly the
same size over a wide range in $S$ and $J$, which seems impossible to
achieve with any reasonably smooth surface energy function.  Instead
of a linear relation, one would instead expect some downward curvature
in Fig.~\ref{scale}, which is consistent with the MD data points
alone, but not in combination with the NPC experiment.

\section{Reconstructing the formation free energy from MD simulations}
\label{sec:free-energy}

We now derive the free energies of cluster formation directly from
 MD results and compare them with predictions from three widely used models:
In the (modified) classical nucleation theory CNT (or MCNT)
and in the semi-phenomenological (SP) model \cite{Dillman1991,Laaksonen1994},
 the free energies $\Delta
G_i$ are respectively
\begin{eqnarray}
\frac{\Delta G_{i,\rm CNT}}{kT} &=& -i \ln S + \eta i^{2/3} \label{CNT},\\ 
\frac{\Delta G_{i,\rm MCNT}}{kT} &=& -(i-1) \ln S + \eta  (i^{2/3}-1) \label{MCNT}, \;\;\textrm{and}\\ 
\frac{\Delta G_{i,\rm SP}}{kT} &=& -(i-1) \ln S + \eta  (i^{2/3}-1)  +  \xi (i^{1/3} -1 )\label{SP} \; , 
\end{eqnarray}
where  $S = P_1 / P_{\rm e}$ is the supersaturation ratio of
monomers using the saturated vapor pressure $P\sub{e}$ and the partial
pressure of monomers $P_1$,  $\eta$ and $\xi$ are
temperature-dependent quantities which can be fixed from the condensed
phase surface tension, bulk density and the second virial coefficient
\cite{Dillman1991,Diemand2013}.
 Note that the CNT assumes large cluster sizes, it
is not expected to work for small clusters and its $\Delta G\sub{i}$
does not vanish at $i=1$, i.e., for monomers.

The formation free energy of a cluster is directly related to the 
equilibrium size distribution 
 $n\sub{e}(i)$: 
\begin{eqnarray}
{\Delta G_i \over kT} = \ln \left( n(1) \over n\sub{e}(i) \right),
\label{delta-g} 
\end{eqnarray} 
where $n(1)$ is the number density of the monomers
\cite{Wedekind2007,Tanaka2011,Diemand2013}.  For small subcritical
clusters ($i \siml i^*$), the steady state size distribution $n(i)$,
which can be measured in MD simulations, agrees very well with the
equilibrium size distribution $n\sub{e}(i)$ \cite{Yasuoka1998}, which
lets us obtain $\Delta G_i$ for small clusters \cite{Yasuoka1998,
Matsubara2007, Tanaka2011, Diemand2013,Tanaka2014}. 
Obtaining the {\it full} free energy landscape, including the crucial
region around the critical sizes, requires a more sophisticated
method, which takes the difference between steady state and
equilibrium size distributions into account.  A first procedure of
this kind was proposed by Wedekind and Reguera\cite{Wedekind2008}
based on mean first passage time (MFPT) method.  In principle it
allows a full reconstruction based on a large number of small
simulations, each one is run until it produces one nucleation event.
However, the observation of one event does not demonstrate that the
simulations are really sampling the assumed steady state nucleation
regime, the passage times might include some initial lag time and a
significant transient nucleation phase, which precedes the steady
state regime \cite{Shneidman1999}. Both time-scales become quite large
for LJ vapor-to-liquid nucleation at low temperatures
\cite{Diemand2013}.

Our recent, very large scale nucleation simulations allow very precise
measurements of the cluster size distribution during a clearly
resolved steady state nucleation regime and under realistic constant
external conditions \cite{Diemand2013}. Here we present a new method to
obtain the full free energy landscape from these steady state size
distributions: The
nucleation rate is the net number of the transition from $i$-mers to
$i+1$-mers and given by
\begin{eqnarray}
J=R^{+}(i)n(i)-R^{-}(i+1)n(i+1),
\label{steadyj} 
\end{eqnarray} 
where $R^{+}(i)$ is the transition rate from a cluster of $i$
molecules, $i$-mer, to ($i$+1)-mer per unit time, {\it i.e.,} the
accretion rate, and $R^{-}(i)$ is the transition rate from $i$-mer to
($i$-1)-mer per unit time, {\it i.e.,} the evaporation rate of
$i$-mer.  $R^{+}(i)$ is given by $ R^+(i) =\alpha n(1) v\sub{th} (4\pi
r_0^2 i^{2/3})$, where $\alpha$ is the sticking probability, 
$v\sub{th}$ is the thermal velocity,
$\sqrt{kT/2 \pi m}$, and $r_0$ is the radius of a monomer, $(3 m / 4
\pi \rho\sub{m})^{1/3}$ where $m$ is the mass of a molecule and
$\rho\sub{m}$ is the bulk density.  The evaporation rate is obtained
from the principle of detailed balance in thermal equilibrium:
\begin{eqnarray}
 R^{-}(i+1)n\sub{e}(i+1)=R^{+}(i)n\sub{e}(i).   
\label{tranm} 
\end{eqnarray}
From Eqs.(\ref{steadyj}) and (\ref{tranm}),  the nucleation rate is given by   
\begin{equation}
J =  
\left[  \sum_{\rm i=1}^{\infty} \frac{1}{R^+(i) n_e(i) }\right]^{-1} \simeq R^+(i_*) n_e(i_*) Z ,
\end{equation}
with  the Zeldovich factor, $Z$. 

\begin{figure}
\includegraphics[height=.2\textheight]{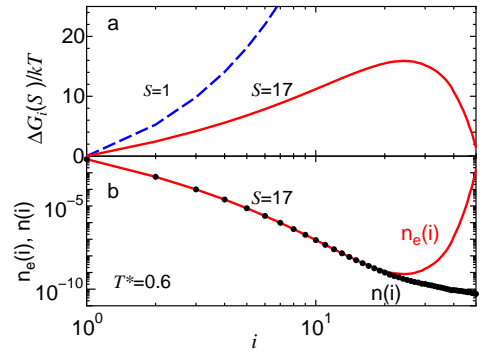}
\caption{(a) $\Delta G_{i}(S) / (kT)$ as a function of $i$, where
 $T^*=kT/\epsilon=0.6$ and $S=16.9$ (T6n80 in Table III in Diemand et
 al. \cite{Diemand2013}).  The dashed line shows $\Delta G_{i}(S) /
 (kT)$ at $S=1$.  (b) The equilibrium number density of $i$-mers
 $n\sub{e}(i)$ [$\sigma^{-3}$] (solid curve) and the steady number
 density obtained by the simulation $n(i)$ [$\sigma^{-3}$] (circles). 
 }\label{neq-deltag}
\end{figure}

From Eqs.(\ref{steadyj}) and (\ref{tranm}), we obtain 
\begin{eqnarray}
{  n\sub{e}(i) \over n(i)  } = { n\sub{e}(i-1)  \over n(i-1) }
\left( 1- {J \over R^{+}(i-1)n(i-1)} \right)^{-1}. 
\label{neq} 
\end{eqnarray}
Equation (\ref{neq}) is a recurrence relation and enables us to obtain
$n\sub{e}(i)$ if $J, n(i)$ and $n\sub{e}(i-1)$ are known
\cite{Tanaka2014}.  Fig.~\ref{neq-deltag} shows $n\sub{e}(i)$, $n(i)$,
and $\Delta G_{i}(S)$ derived by Eq.~(\ref{neq}) for a typical example
($T^*=kT/ \epsilon=0.6$ and $S=16.9$ which corresponds to the case
T6n80 in Table III in Diemand et al. \cite{Diemand2013}).  $\Delta
G_{i}(S\!=\!1)$ is a surface term corresponding to the work required to
form the vapor-liquid interface.  From Eq.~(\ref{neq}), we obtain
$\Delta G_i(S\!=\!1)$:
\begin{eqnarray}
\Delta G_i(S\!=\!1)=\Delta G_i(S) 
 +(i-1) \ln S, 
\end{eqnarray}
using the dependence of the supersaturation in the theories 
except the CNT. Fig.~\ref{neq-deltag}  also shows $\Delta G_i(S\!=\!1)$. 

\begin{figure}
\includegraphics[height=.4\textheight]{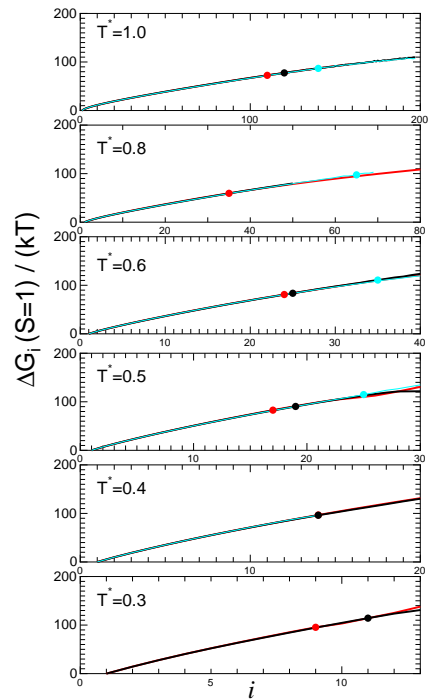}
\caption{$\Delta G_{i}(S\!=\!1)$ as a function of $i$ for various
 temperatures. At each temperature, we show $\Delta G_{i}(S\!=\!1)$
 obtained by the different values of the supersaturation ratio.  The
 circles show the critical clusters derived by the maximum of $\Delta
 G_{i}(S)$ for various supersaturation ratios $S$.
  We can confirm $\Delta G_{i}(S\!=\!1)$ depends on only $T$.
 }\label{deltag-s1}
\end{figure}

The surface terms of free energy $\Delta G_{i}(S\!=\!1)$ at various
temperatures and supersaturation ratios obtained by MD simulations are
shown in Figure~\ref{deltag-s1}, where we evaluated $R^{+}(i)$ using
$\alpha$ obtained by the MD simulations (Table III in Diemand et al.).
From Figure~\ref{deltag-s1}, we confirm $\Delta G_{i}(S\!=\!1)$ depends
only on temperature, which implies that the volume term in Eqs.(\ref{MCNT})
 and (\ref{SP})  works very well. 

The peaks of the free energy curves 
give critical cluster sizes which
agree very well with those from the nucleation theorem (see
Fig. \ref{scale2}).  Since the nucleation rates, which enter into the
nucleation theorem, do not depend on the detailed cluster definition,
this good agreement provides a robust confirmation, that the simple
Stillinger criterion used here\cite{Tanaka2011,Diemand2013} gives
realistic cluster size estimates.
An earlier study\cite{Wedekind2007} found that
critical sizes based on the Stillinger definition are up to a factor 2
larger than independent estimates from the nucleation theorem. This
contradiction can be resolved by a detailed comparison with other MD
simulations at very similar conditions\cite{Tanaka2011}: Using the
initial supersaturations $S_0$ in the nucleation theorem (as in
\cite{Wedekind2007}) instead of the actual supersaturation
$S$ during the simulation\cite{Tanaka2011}, leads one to underestimate
the critical sizes by up to a factor of 1.8, which eliminates the
discrepancy reported in \cite{Wedekind2007}.

\begin{figure}
\includegraphics[height=.35\textheight]{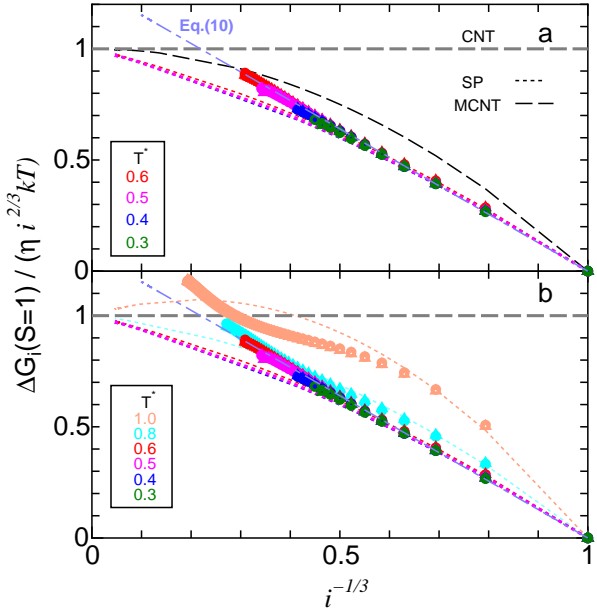}
\caption{(a) $\Delta G_{i}(S\!\!=\!\!1)/(\eta i^{2/3} kT)$ as a function of
 $i^{-1/3}$ at $kT/\epsilon \le 0.6$.  Results obtained from 11 MD
 simulations are plotted with  symbols: different symbols indicate
 the MD results starting from different supersaturation ratios.  We
 find that they are universal, which implies that $\Delta
 G_{i}(S\!\!=\!\!1)/(\eta i^{2/3} kT)$ is independent of temperature for $(T
 \le 0.6)$.  From the fitting, we obtain $\Delta G_{i}(S\!=\!1)/(\eta
 i^{2/3} kT)= 1.28(1-i^{-1/3})$ (the dotted-dashed line).  The results
 by the SP (dotted lines) and MCNT (dashed line) are also shown.
 $\Delta G_{i}(S\!=\!1)/(\eta i^{2/3} kT)=1$ in the CNT.  (b) The same as
 (a) but for all temperatures.  }\label{deltag}
\end{figure}

\begin{figure}
\includegraphics[height=.25\textheight]{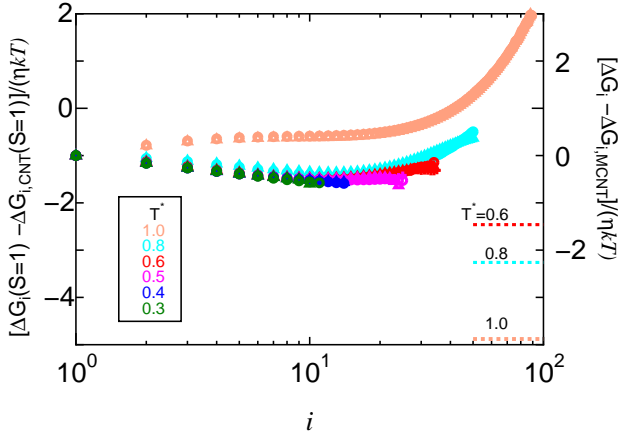}
\caption{  The difference in $\Delta G_{i}(S\!\!=\!\!1)$ between MD
results and the CNT divided by $\eta kT$, $[\Delta G_{i}(S\!=\!1)-\Delta
G_{i, \rm CNT}(S\!=\!1)]/(\eta kT)$ as a function of $i$ at various
temperatures.  
Different symbols indicate the MD results starting from
different supersaturation ratios. 
The results of McGraw and Laaksonen\cite{McGraw1997} are also
shown by dotted lines for $T^*=0.6, 0.8$ and 1.0. 
 The right vertical axis shows the value of the 
difference between the MD results and the MCNT, $[\Delta G_{i}-\Delta
G_{i, \rm MCNT}]/(\eta kT)$, which  is valid for any value of $S$. 
}\label{deltag-diff}
\end{figure}

\section{A new scaling for nucleation rates}\label{sec:new-scaling}

Fig.~\ref{deltag} shows the surface energy $\Delta G_{i}(S\!=\!1)$ divided
by that of the CNT, $\Delta G_{i}(S\!=\!1)/(\eta i^{2/3} kT)$, as a
function of $i^{-1/3}$.  The theoretical evaluations are also shown in
Fig.~\ref{deltag}.  The simulation results agree with the SP model at
$0.5 \siml i^{-1/3}<1$, but deviate from the model for larger clusters
of $i^{-1/3}<0.5$.  Surprisingly, $\Delta G_{i}(S\!=\!1)/(\eta i^{2/3}
kT)$ is almost the same for all results obtained by 11 MD simulations
for temperatures below the triple point.  This indicates that
$\Delta G_{i}(S\!=\!1)/(\eta i^{2/3} kT)$ is a function of $i$ and
independent of the temperature.  From the fitting of the results, we
obtain
\begin{eqnarray}
{\Delta G_{i}(S\!=\!1) \over \eta
 i^{2/3} kT }= f(i)=A(1-i^{-1/3}), 
\label{fit}
\end{eqnarray}
where $A=1.28$. The fitting function is also shown by the
dotted-dashed line in Fig.~\ref{deltag}.  Equation~(\ref{fit}) implies
a constant, positive Tolman length of $\delta = 0.5 r_0$ and the
constant $A$ sets an effective normalisation factor for the planar
surface energy (or the surface area), if we interpret $\Delta
G_{i}(S\!=\!1)/ (\eta i^{2/3} kT )= a_i \gamma_i/ (4 \pi r_0^2
\gamma)$, where $\gamma_i = \gamma [1-2\delta/(r_0 i^{1/3})]$ and
$a_i$ are the surface tension and surface area of the cluster and
$\gamma$ is the planar surface tension.  Equation~(\ref{fit}) could be
a promising candidate for an accurate nucleation theory, in which $A$
is temperature independent below the triple point.  Our result
indicates that at low temperatures the Tolman relation is valid even
for very small clusters including 2-30 atoms.

 McGraw and Laaksonen\cite{McGraw1996,McGraw1997} obtained $\Delta G_i$
 of large clusters $(i \simg 50)$ with 
 density functional calculations. 
 They found that the deviation of $\Delta G_i$ from the CNT 
 is temperature dependent, but independent of the cluster size.
 Figure~\ref{deltag-diff} shows the difference of $\Delta G_{i}(S\!=\!1)$
 between MD results and the CNT, {\it i.e.,}
 $\Delta G_{i}(S\!=\!1)-\Delta G_{i, \rm CNT}(S\!=\!1)$ as a function
 of $i$.  We find these differences are nearly constant 
 around $i \sim 10$ for
 each temperature.  But they 
 increase with the size for $i > 20$.
 According to McGraw and Laaksonen (1997)\cite{McGraw1997}, on the other hand,
 $[\Delta
 G_{i}(S\!=\!1)-\Delta G_{i,\rm CNT}(S\!=\!1)]/(\eta kT)$ are calculated to be
 -2.46, -3.26, and -4.88 for $T^*=0.6, 0.8$, and 1.0, respectively.

Using Eq.~(\ref{fit}), 
the  critical cluster $i_*$ is obtained by 
\begin{eqnarray}
i_*&=& \left( A \eta  \over 3 \ln S \right)^3
\left( 1 + \sqrt{1-{3 \over A}{\ln S \over \eta}}
    \right)^{3},
\label{criticalstar}
\end{eqnarray}
from the following relation 
\begin{eqnarray}
 - {\ln S  \over \eta} +
 { 2 \over 3} i_*^{-1/3} f(i*)+ i_*^{2/3} f'(i_*)=0, 
\label{condition}
\end{eqnarray}
 where we have assumed that the molecular volume is far smaller in the liquid phase than in the gas phase. The detailed derivation is given in the Appendix.

\begin{figure}
\includegraphics[height=.2\textheight]{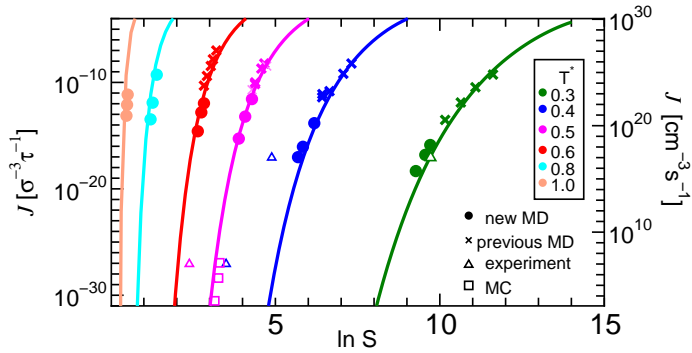}
\caption{The nucleation rate as a function of the supersaturation ratio.
  The analytical formula for the nucleation rates are shown by solid
  lines.  The results for various temperature and supersaturation
  ratios by the large-scale MD simulations \cite{Diemand2013} and the
  previous ones \cite{Wedekind2007,Tanaka2011} 
 are shown by the filled circles and the
  crosses, respectively. The results for MC simulations\cite{Hale2010}
  are shown by the squares, where the temperature is $T^*=0.5$.  The
  triangles show the experimental results for argon
  \cite{Iland2007,Sinha2010}. }\label{scale1}
\end{figure}

We also derive the analytical formula for the nucleation rate:
\begin{eqnarray}
\ln J'  
 &=& \ln [\alpha Z i_*^{2/3} ] +(i_*+1) \ln S  -
 i_*^{2/3} \eta f(i_*) ,
\label{rate3}
\end{eqnarray}
where $J'$ is  a dimensionless nucleation rate defined by 
$J'=J / (4 \pi r_0^2 n\sub{sat}^2 v\sub{th} )$ with
 the saturated  number density of monomers $n\sub{sat}$ $(=n(1)/S)$
 and    the Zeldvich factor is  given by 
\begin{eqnarray}
Z=  {1 \over 3} i_*^{-2/3}
   \sqrt{ {A \eta \over \pi} (1-i_*^{-1/3})}. 
\label{zeldvich}
\end{eqnarray}
Fig.~\ref{scale1} shows the nucleation rate as a function of $\ln S$
obtained by the MD simulations and the analytical formula.  We find
good agreements between the analyses and the simulations for the
various temperatures and supersaturation ratios. 

\begin{figure}
\includegraphics[height=.3\textheight]{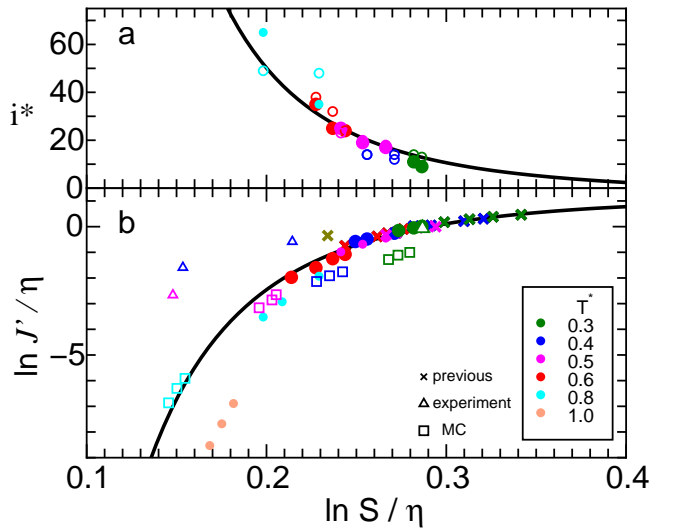}
\caption{ We propose that (a) the size of critical cluster and (b) $
 \ln J'/\eta$ are determined only by $\ln S / \eta$, where $J'=J/(4
 \pi r_0^2 n\sub{sat}^2 v\sub{th})$.  The analytical formula obtained
 by our model are shown by the solid lines.  Panel (a) shows, that the
 critical clusters sizes derived from the maximum of $\Delta G_{i}(S)$
 (filled circles) and from the nucleation theorem (open circles,
 $i\sub{NT}$ in Diemand et al. \cite{Diemand2013}) agree very well
 with each other and also with our analytical model.  In (b), the
 results for various temperature and supersaturation ratios by the
 large-scale MD simulations \cite{Diemand2013} and the previous ones
 \cite{Yasuoka1998,Wedekind2007,Tanaka2011} are shown by the filled
 circles and the crosses, respectively.  The results for MC
 simulations\cite{Hale2010} are shown with square markers . The triangles
 show the experimental results for argon \cite{Iland2007,Sinha2010}.
 }\label{scale2}
\end{figure}

Our finding that $\Delta G_{i}(S\!=\!1)/(\eta i^{2/3} kT)$ is independent
of the temperature leads to a scaling relation.
Equation~(\ref{condition}) indicates that $i_*$ is a function of only
$\ln S / \eta$.  Thus from Eq.~(\ref{rate3}) $\ln J'/\eta$ is
determined only by $\ln S / \eta$, neglecting a term including
Zeldovich factor which is smaller than the other terms.
Fig.~\ref{scale2} shows the size of critical clusters and $\ln J'
/\eta$ obtained by MD and MC simulations and experiments as a function of
$\ln S / \eta$.  We confirm that $\ln J'/\eta$ is scaled by $\ln S
/ \eta$ almost perfectly for MD simulations, at $T^* \le 0.6$.  At
high temperatures ($T^*>0.8$), $\ln J'/\eta$ deviates from the scaling
relation.  This would come from the deviation of $f(i)$, i.e., $f(i)$
depends on $T$ at $T^*>0.8$ (see Fig.~\ref{deltag}(b)).
Fig.~\ref{scale2} shows this scaling also works for one SNN experiment
($T^*=0.3$) and the MC simulations at $T^*=0.5$ and 0.7, although 
some MC data and experiments deviate from the scaling relation.

\section{Summary and conclusions}

 We derived for the first time the formation free energy of a cluster
 over a wide range of cluster sizes and temperatures from recent very
 large-scale MD simulations. The peaks of the free energy curves give
 critical cluster sizes, which agree well with independent estimates
 based on the nucleation theorem.  This implies that the simple
 Stillinger criterion used here gives realistic cluster size
 estimates.
 
 At low temperatures the free energies show a universal deviation from
 the CNT, which allows us to derive a new scaling relation for
 nucleation: $\ln J' / \eta$ is scaled by $\ln S / \eta$. This scaling
 relation predicts the critical cluster size very well. The relation
 can be explained from a surface energy required to form the
 vapor-liquid interface and implies a constant, positive Tolman length
 of $\delta = 0.5 r_0$.  Generally, $\Delta G_i(S\!=\!1)$ is written as
 the surface energy multiplied by the surface area, $a_i \gamma_i$.
 In the theory, the cluster is always assumed to be spherical and has
 the same density as the bulk liquid. However, our analyses of cluster
 properties show larger surface areas (Ang\'elil et
 al. \cite{Angelil2014}). The higher normalisation ($A \simeq 1.28$ in
 Eq.~(\ref{fit})) of $\Delta G_i(S)$ relative to the models might be
 caused by these larger surface areas. The scaling relation and the
 relation between the cluster properties and $\Delta G_i$ should be
 investigated in more detail for various materials.



\section{Acknowledgments}

 We thank the
 anonymous reviewers for their valuable suggestions which have improved the
 quality of the paper.  This work was supported by the Japan Society
 for the Promotion of Science (JSPS). J.D. and R.A. acknowledge support from the Swiss National Science Foundation (SNSF).

\section{Appendix}

The general expression for the minimum work $\Delta G (r)$ required to form a cluster 
 of radius $r$, 
is given by\cite{landau}
\begin{eqnarray}
\Delta G (r) =  {V\sub{l} \over v\sub{l}}
 [\mu\sub{l} (P\sub{l}) - \mu\sub{g} (P\sub{g})]
 - (P\sub{l}-P\sub{g}) V\sub{l} + a_i \gamma_i, 
\end{eqnarray}
where $\mu\sub{l}$ and  $\mu\sub{g}$ are 
 the chemical potentials of liquid and gas,
 $P\sub{l}$ and $P\sub{g}$ are the pressures of metastable liquid
 and gas, and  $v\sub{l}$ and  $V\sub{l} (=i v\sub{l} = 4 \pi r^3/ 3)$
 are the molecular volumes  of liquid  and
 the volumes of a cluster respectively. 
Using $\mu\sub{l}(P\sub{e})=\mu\sub{g}(P\sub{e})$ and 
 $\mu\sub{l}(P\sub{l})-\mu\sub{l}(P\sub{e})=v\sub{l}(P\sub{l}-P\sub{e})$, 
we obtain 
\begin{eqnarray}
\Delta G (r) &=&  {V\sub{l} \over v\sub{l}}
 [\mu\sub{g} (P\sub{e}) + v\sub{l} (P\sub{l}-P\sub{e})
- \mu\sub{g} (P\sub{g})] \nonumber \\
& &  - (P\sub{l}-P\sub{g}) V\sub{l} + a_i \gamma_i,  \nonumber \\
&=&  {V\sub{l} \over v\sub{l}}
[\mu\sub{g} (P\sub{e}) - \mu\sub{g} (P\sub{g})]
 + (P\sub{g}-P\sub{e}) V\sub{l} + a_i \gamma_i,  \nonumber \\
&=&  {V\sub{l} \over v\sub{l}}
[\mu\sub{g} (P\sub{e}) - \mu\sub{g} (P\sub{g})]
 + i (S-1) kT  { v\sub{l} \over v\sub{g}} + a_i \gamma_i,  \nonumber \\
\label{app1}
\end{eqnarray}
where $v\sub{g}$ is the molecular volume in the gas phase.
For the case  $v\sub{l} << v\sub{g}$, 
 the second term on the right hand side of Eq.~(\ref{app1})
is negligible. Assuming $a_i \gamma_i=4 \pi r^2 \gamma$,
 we obtain the formula for the critical radius $r\sub{cr}$ 
 called the Kelvin relation\cite{Laaksonen1996}  
from $\partial \Delta G (r)/ \partial r =0$:
\begin{eqnarray}
r\sub{cr} = { 2 \gamma v\sub{l} \over 
  \Delta \mu }, 
\label{app2}
\end{eqnarray}
where $\Delta \mu = \mu (P\sub{g}) - \mu\sub{g} (P\sub{e})=kT \ln S$.

From Eq.~(\ref{fit}), 
the result from MD simulations shows 
\begin{eqnarray}
a_i \gamma_i=4 \pi r^2 A \gamma (1-r_0/r),
\label{app3}
\end{eqnarray}
thus we obtain the following relation at $r=r\sub{cr}$:
\begin{eqnarray}
 - {4 \pi r\sub{cr}^2  \Delta \mu \over v\sub{l}}
 + 8 \pi r\sub{cr} A \gamma - 4 \pi r_0 A \gamma =0.
\label{app4}
\end{eqnarray}
From Eq.~(\ref{app4}), $r\sub{cr}$ is given by
\begin{eqnarray}
 r\sub{cr} = {A \gamma v\sub{l} \over  \Delta \mu} 
\left(  1 + \sqrt{ 1 - {\Delta \mu r_0 \over v\sub{l} A  \gamma }}
 \right),
\label{app5}
\end{eqnarray}
which corresponds to Eq.~(\ref{criticalstar}).

\bibliography{jcp.bib}   

\begin{thebibliography}{9}

\bibitem{Volmer1926}
M. Volmer and A. Weber, Z. Phys. Chem. {\bf 119}, 277 (1926).

\bibitem{Becker1935}
V. R. Becker and W. D\"{o}ring, Ann. Phys. {\bf 416}, 719 (1935).

\bibitem{Feder1966}
J. Feder, K. C. Russell, J. Lothe, and G. M. Pound, Adv. Phys. {\bf 15},
 111 (1966).

\bibitem{Anderson1980}
R. J. Anderson, R. C. Miller, J. L. Kassner, and D. E. Hagen, 
J. Atmos. Sci. {\bf 37}, 2508 (1980).

\bibitem{Schmitt1982}
J. L. Schmitt, G. W. Adams, and R. A. Zalabsky, J. Chem. Phys. {\bf 77}, 2089
(1982).

\bibitem{Schmitt1983}
J. L. Schmitt, R. A. Zalabsky, and G. W. Adams, J. Chem. Phys. {\bf 79}, 4496
(1983).

\bibitem{Adams1984}
G. W. Adams, J. L. Schmitt, and R. A. Zalabsky, J. Chem. Phys. {\bf 81}, 5074
(1984).

\bibitem{Dillman1991}
A. Dillman and G. E. A. Meier,  J. Chem. Phys. {\bf 94}, 3872 (1991).

\bibitem{Oxtoby1992}
D. W. Oxtoby, J. Phys.: Condens. Matter {\bf 4}, 7627 (1992).

\bibitem{Viisanen1993}
Y. Viisanen, R. Strey, and H. Reiss, J. Chem. Phys. {\bf 99}, 4680 (1993).

\bibitem{Laaksonen1994}
A. Laaksonen, I. J. Ford, and M. Kulmala, Phys. Rev. E {\bf 49}, 5517 (1994).

\bibitem{Viisanen1994}
Y. Viisanen and R. Strey, J. Chem. Phys. {\bf 101}, 7835 (1994).

\bibitem{Hameri1996}
K. H\"{a}meri and M. Kulmala, J. Chem. Phys. {\bf 105}, 7696 (1996).




\bibitem{Iland2007}
K. Iland, J. W\"olk, R. Strey,  and D. Kashchiev,
 J. Chem. Phys. {\bf 127}, 154506 (2007).

\bibitem{Sinha2010}
S. Sinha, A. Bhabhe, H. Laksmono, J. W\"{o}lk, R. Strey, and B. Wyslouzil,
J. Chem. Phys. {\bf 132}, 064304
(2010).

\bibitem{Yasuoka1998}
K. Yasuoka and M. Matsumoto, J. Chem. Phys. {\bf 109}, 8451 (1998).

\bibitem{Yasuoka1998b}
K. Yasuoka and M. Matsumoto, J. Chem. Phys. {\bf 109}, 8463 (1998).

\bibitem{Wolde1998}
P. R. ten Wolde and D. Frenkel, J. Chem. Phys. {\bf 109}, 9901 (1998).


\bibitem{Oh1999}
K. J. Oh and X. C. Zeng, J. Chem. Phys. {\bf 110}, 4471 (1999).

\bibitem{Senger1999}
B. Senger, P. Schaaf, D. S. Corti, R. Bowles, D. Pointu, J.-C. Voegel, and
H. Reiss, J. Chem. Phys. {\bf 110}, 6438 (1999).

\bibitem{Wolde1999}
P. R. ten Wolde, M. J. Ruiz-Montero, and D. Frenkel, J. Chem. Phys. {\bf 110},
1591 (1999).

\bibitem{Laasonen2000}
K. Laasonen, S.Wonczak, R. Strey, and A. Laaksonen, J. Chem. Phys. {\bf 113},
9741 (2000).

\bibitem{Oh2000}
K. J. Oh and X. C. Zeng, J. Chem. Phys. {\bf 112}, 294 (2000).


\bibitem{Vehkamaki2000}
H. Vehkam\"{a}ki and I. J. Ford, J. Chem. Phys. {\bf 112}, 4193 (2000).

\bibitem{Chen2001}
B. Chen, J. I. Siepmann, K. J. Oh, and M. L. Klein, J. Chem. Phys. {\bf 115},
10903 (2001).

\bibitem{Schaaf2001}
P. Schaaf, B. Senger, J.-C. Voegel, R. K. Bowles, and H. Reiss, J. Chem.
Phys. {\bf 114}, 8091 (2001).

\bibitem{Wold2001}
J. W\"{o}ld and R. Strey, J. Phys. Chem. B {\bf 105}, 11683, (2001). 





\bibitem{merikanto2004}
J. Merikanto, H. Vehk\"{a}maki, and E. Zapadinsky,
J. Chem. Phys. {\bf 121}, 914 (2004).

\bibitem{Tanaka2005}
K. K. Tanaka, K. Kawamura, H. Tanaka, and  K. Nakazawa, J. Chem. Phys. 
{\bf 122}, 184514 (2005).

\bibitem{Matsubara2007}
H. Matsubara, T. Koishi, T. Ebisuzaki, and K. Yasuoka, J. Chem. Phys. 
{\bf 127}, 214507 (2007).




\bibitem{Wedekind2007}
J. Wedekind, J. W\"{o}lk,  D. Reguera,
 and R. Strey, J. Chem. Phys. {\bf 127}, 154515 (2007).

\bibitem{Horsch2008}
M. Horsch, J. Vrabec, and H. Hasse, 
Phys. Rev. E {\bf 78}, 011603 (2008).

\bibitem{Horsch2009}
M. Horsch,  and J. Vrabec,
J. Chem. Phys. {\bf 131}, 184104 (2009).

\bibitem{Wedekind2009}
J. Wedekind, G. Chkonia, J. W\"{o}lk, R. Strey, and D. Reguera, 
J. Chem. Phys. {\bf 131}, 114506 (2009).

\bibitem{Napari2010}
I. Napari, J. Julin, and H. Vehkam\"{a}ki, 
J. Chem. Phys. {\bf 133}, 154503 (2010).

\bibitem{Tanaka2011}
K. K. Tanaka, H. Tanaka, 
T. Yamamoto, and K. Kawamura, J. Chem. Phys. {\bf 134}, 204313, (2011).


\bibitem{Diemand2013}
J. Diemand, R. Ang\'elil, 
K. K. Tanaka, and H. Tanaka, J. Chem. Phys. {\bf 139}, 074309. (2013).

\bibitem{Tanaka2014}
 K. K. Tanaka, A. Kawano, and  H. Tanaka, J. Chem. Phys.
 {\bf 140}, 114302 (2014).

\bibitem{Hale1986}
B. N. Hale, Phys. Rev. A {\bf 33}, 4156 (1986).

\bibitem{Hale1992}
B. N. Hale, Metall. Trans. A {\bf 23A}, 1863 (1992). 

\bibitem{Hale2005}
B. N. Hale, J. Chem. Phys. {\bf 122}, 204509 (2005). 

\bibitem{Hale2010}
B. N. Hale and M. Thomason, Phys. Rev. Let. {\bf 105}, 046101 (2010).

\bibitem{Kalikmanov}
 V. I. Kalikmanov, {\it Nucleation theory}, Lecture notes in Physics
  (Springer, Dordrecht, 2013) , Vol. 860. 

\bibitem{Wedekind2008}
J. Wedekind  and D. Reguera, J. Phys. Chem. B  {\bf 112},
11060 (2008).

\bibitem{Shneidman1999}
V. Shneidman, K. Jackson, and
K. Beatty, Phys. Rev. B {\bf 59}, 3579 (1999)


\bibitem{McGraw1996}
R. McGraw and A. Laaksonen, Phys. Rev. Lett. {\bf 76}, 2754 (1996).

\bibitem{McGraw1997}
R. McGraw and A. Laaksonen, J. Chem. Phys. {\bf 106}, 5284 (1997).

\bibitem{Angelil2014}
 R. Ang\'elil, J. Diemand, K. K. Tanaka, and H. Tanaka, J. Chem. Phys.
{\bf 140}, 074303 (2014).

\bibitem{landau}
L. D. Landau and E. M. Lifshitz, 
 {\it Stastistical Physics} (Pergamon press, Oxford, 1980), section 162. 

\bibitem{Laaksonen1996}
 A. Laaksonen and R. McGraw, Europhys. Lett. {\bf 35}, 367 (1996).


\end{thebibliography}
\bibliographystyle{aipprocl} 


\end{document}